\documentclass[a4paper,12pt,reqno]{amsart}

\usepackage{a4}
\usepackage{verbatim}
\usepackage{todonotes}
\usepackage{color}

\usepackage{amssymb}
\usepackage{color}
\usepackage{graphicx}
\usepackage{floatflt}
\usepackage{float} 		%damit H funktioniert bei bildern
\usepackage{amsthm}
\usepackage{amscd}
\usepackage{hyperref}

\newtheorem{definition}{Definition}
\newtheorem{example}[definition]{Example}

\newtheorem{theorem}[definition]{Theorem}
\newtheorem{lemma}[definition]{Lemma}
\newtheorem{remark}[definition]{Remark}

\newtheorem{corollary}[definition]{Corollary}
\newtheorem{corollary2}[definition]{*Corollary}

\newtheorem{proposition}[definition]{Proposition}

\def\XXint#1#2#3{{\setbox0=\hbox{$#1{#2#3}{\int}$}
		\vcenter{\hbox{$#2#3$}}\kern-.5\wd0}}

\makeatletter
\@addtoreset{definition}{section}
\@addtoreset{equation}{section}
\makeatother

\newcommand{\G}{\mathcal{G}}

\newcommand{\I}{\mathcal{I}}

\DeclareMathOperator{\Res}{Res}

\allowdisplaybreaks[4]

\begin{document}
%\pagenumbering{arabic}

\title[Laplace transform of the $x-y$ formula]{Laplace transform of the $x-y$ symplectic transformation formula in Topological Recursion}

\author{Alexander Hock}

\address{Mathematical Institute, University of Oxford, Andrew Wiles Building, Woodstock Road,
	OX2 6GG, Oxford, UK 
{\itshape email address:} \normalfont  
\texttt{alexander.hock@maths.ox.ac.uk}}

\maketitle

\markboth{\hfill\textsc\shortauthors}{\textsc{Laplace transform of the $x-y$ formula}\hfill}

%{\footnotesize\tableofcontents}

\begin{abstract}
	 The functional relation coming from the $x-y$ symplectic transformation of Topological Recursion has a lot of applications, for instance it is the higher order moment-cumulant relation in free probability or can be used to compute intersection numbers on the moduli space of complex curves. We derive the Laplace transform of this functional relation, which has a very nice and compact form as a formal power series in $\hbar$. We apply the Laplace transformed formula to the Airy curve and the Lambert curve. 
\end{abstract}

\section{Introduction}
Topological Recursion (TR) is a universal structure which generates from the so-called spectral curve a family of multi-differentials $\omega_{g,n}$ on the spectral curve (Riemann surface). TR occurs to be related to seemingly different areas of mathematics and mathematical physics. To give an incomplete list, it is related to volumes of moduli spaces, Hurwitz numbers, intersection numbers of moduli spaces, Gromov-Witten theory, enumerative combinatorics, random matrix theory, quantum field theory on noncommutative spaces, free probability and quantum knot theory \cite{Eynard:2014zxa}. 

Knowing properties which hold in general for any spectral curve can give therefore new insight into the applications of TR. For instance, the multi-differential $\omega_{g,n}$ are symmetric, but generated via a non-symmetric formula. This symmetry is in almost all application obvious from the beginning. Transforming the spectral curve under a \textit{symplectic transformation} can leave the $\omega_{g,n}$ invariant. From this one can deduce that certain models for instance in random matrix theory are equivalent. However, there is a very specific symplectic transformation the \textit{$x-y$ symplectic transformation} which leaves actually the spectral curve invariant but generates completely different multi-differential. Recently, the relation between these two different families of multi-differentials were found in its most simple representation \cite{Hock:2022pbw,Alexandrov:2022ydc}. The result is a functional relation which has already generalised the higher order moment-cumulant relation in free probability \cite{Borot:2021thu}. Equivalently, this functional relation relates fully simple and ordinary maps in enumerative combinatorics \cite{Borot:2021eif,Bychkov:2021hfh}. Due to free probability, it can be understood as the qunatised version of a moment-cumulant relation.
The $x-y$ symplectic transformation also reproves known results for intersection numbers on the moduli space of complex curves $\overline{\mathcal{M}}_{g,n}$ and might give new algorithms or closed formulas to compute them. 

An other important tool in TR is the Laplace transformation. In \cite{Eynard:2011kk,Eynard2011InvariantsOS}, it was shown that the Laplace transform of the $\omega_{g.n}$ has a direct interpretation in terms of intersection numbers on $\overline{\mathcal{M}}_{g,n}$. As an application of TR in Gromov-Witten theory, or more generally, in topological string theory, the Laplace transform has an interpretation as \textit{mirror symmetry}. More precisely, the $A$-model and $B$-model are two different approaches to studying the geometry of Calabi-Yau manifolds in topological string theory. They are related through mirror symmetry. Based on observations in \cite{Bouchard:2007ys}, the $A$-model and $B$-model are related to TR and the mirror map has the interpretation of the Laplace transform (see also \cite{Eynard2009TheLT,Mulase:2012yk} for details).

Consequently, it is completely natural to bring together the functional relation of the $x-y$ symplectic transformation and the Laplace transform. Actually, we observe that the functional relation behaves very well under the Laplace transformation. The differential operator in the functional relation becomes after Laplace transform a multiplication which sums up perfectly in terms of formal power series. The Laplace transform of $\omega_{g,n}$ and of its not-necessarily connected sibling is given in Corollary \ref{cor:LaplaceConW} and Proposition \ref{Pro:Laplace}.

The functional relation is not valid in general for spectral curves with logarithmic singularities. However, we consider the example of the Lambert curve \cite{Bouchard:2007hi}, which encodes Hurwitz numbers. After a small transformation of this curve, we apply successfully the Laplace transform formula of the $x-y$ symplectic transformation to compute Hurwitz numbers. This is the first step into generalising the $x-y$ symplectic transformation to spectral curves with logarithmic singularities, which will have application in topological string theory and quantum knot theory.

\subsection*{Acknowledgement}
This work was supported through
the Walter-Benjamin fellowship\footnote{``Funded by
	the Deutsche Forschungsgemeinschaft (DFG, German Research
	Foundation) -- Project-ID 465029630}.

\section{Formula for $x-y$ Symplectic Transformation in Topological Recursion}\label{Sec:TR}
We start to recap the theory of TR and its property under $x-y$ symplectic transformation in more details.
TR is an algorithm which computes recursively in the negative Euler characteristic $-\chi=2g+n-2$ from some initial data, the so-called \textit{spectral curve}, a family of multi-differentials denoted by $\omega_{g,n}$, which are also commonly called \textit{correlators}.

More precisely, the spectral curve is the tuple $(\Sigma,x,y,B)$, where $\Sigma$ is a compact Riemann surface with $x,y:\Sigma\to \mathbb{C}$ are meromorphic functions with simple and distinct ramification points on $\Sigma$. The multi-differentials $\omega_{g,n}$ live on $\Sigma^n$ with $\omega_{0,1}(z)=y(z)\,dx(z)$ and $\omega_{0,2}=B$, where $B$ is symmetric with double pole on the diagonal and no residue, bi-residue 1 and normalised such that the $A$-periods vanish. In particular for $\Sigma=\mathbb{P}^1$ the complex projective line (Riemann sphere), the bilinear differential is $B(z_1,z_2)=\frac{dz_1\,dz_2}{(z_2-z_2)^2}$.

Then for negative Euler characteristic $\chi<0$, all $\omega_{g,n}$ are defined via \cite{Eynard:2007kz}
\begin{align}
  \label{eq:TR-intro}
&  \omega_{g,n+1}(I,z)
  \\
  & :=\sum_{\beta_i}
  \Res\displaylimits_{q\to \beta_i}
  K_i(z,q)\bigg(
  \omega_{g-1,n+2}(I, q,\sigma_i(q))
  +\!\!\!\!\!\!\!\!
   \sum_{\substack{g_1+g_2=g\\ I_1\uplus I_2=I\\
            (g_i,I_i)\neq (0,\emptyset)}}\!\!\!\!\!\!\!
   \omega_{g_1,|I_1|+1}(I_1,q)
  \omega_{g_2,|I_2|+1}(I_2,\sigma_i(q))\!\bigg). \nonumber
\end{align}
The following notation is used:
\begin{itemize}
\item  $I=\{z_1,\dots,z_n\}$ is a collection of $n$ variables $z_j$
\item  the ramification points
$\beta_i$ of $x$ are defined by $dx(\beta_i)=0$
\item  the local Galois involution $\sigma_i\neq \mathrm{id}$ with
$x(q)=x(\sigma_i(q))$ is defined in the vicinity of $\beta_i$ with fixed point $\beta_i$
\item the recursion kernel $K_i(z,q)$ is also locally 
defined in the vicinity of $\beta_i$ by
\begin{align*}
K_i(z,q)=\frac{\frac{1}{2}\int^{q}_{\sigma_i(q)}
  B(z,\bullet)}{\omega_{0,1}(q)-\omega_{0,1}(\sigma_i(q))}.
\end{align*}
\end{itemize}
It is not obvious from the definition, but 
it turns out that all $\omega_{g,n}$ are symmetric in its variables, and for $\chi<0$ all
 $\omega_{g,n}$ have poles just located at the ramification points, with vanishing residues. In other words, for $\chi<0$ the $\omega_{g,n}$ are exact 1-forms in each variable. We define the primitives
 \begin{align}\label{Phign}
     \Phi_{g,n}(x_1(z_1),...,x_n(z_n)):=\int_o^{z_1}...\int_o^{z_n}\omega_{g,n}(z_1,...,z_n)
 \end{align}
with base point $o$ which will not play any role. Note that $\Phi_{0,1}$ and $\Phi_{0,2}$ might not be globally defined on $\Sigma$, since they may have some branch cuts.

The special case $n=0$ is usually denoted by $\mathcal{F}^{(g)}=\omega_{g,0}$, the so-called \textit{free energy} of genus $g$. They can be obtained by $\Phi_{0,1}(z)$. For negative Euler characteristic $\chi<0$, the free energies are defined by:
\begin{align}
\label{dilaton2}
\sum_{\beta_i} \Res_{z \to \beta_i} \Phi_{0,1}(z)\omega_{g,1}(z)=(2-2g)\mathcal{F}^{(g)} \qquad g>1.
\end{align}
The free energies $\mathcal{F}^{(0)}$ and $\mathcal{F}^{(1)}$ obey a more involved formula, see  \cite{Eynard:2007kz}.

For later purpose, we define also $W_{g,n}$ via
\begin{align}
    W_{g,n}(x_1(z_1),...,x_n(z_n))dx_1(z_1)...dx_n(z_n):=\omega_{g,n}(z_1,...,z_n).
\end{align}
The genus summation with formal parameter $\hbar$ will be denoted by 
\begin{align}\label{Wn}
    W_n(x_1,...,x_n):=&\sum_{g=0}^\infty \hbar^{2g+n-2}W_{g,n}(x_1,...,x_n)\\ \label{Phin}
    \Phi_n(x_1,...,x_n):=&\sum_{g=0}^\infty \hbar^{2g+n-2}\Phi_{g,n}(x_1,...,x_n).
\end{align}

\begin{example}\label{Ex:1}
For $\Sigma=\mathbb{P}^1$ and $x$ unramified, i.e. $x$ has no ramification point, then all $W_{g,n}=0$ for $\chi=2g+n-2<0$. The correlators with positive Euler characteristic are
\begin{align*}
    W_{0,1}(x(z))=&y(z),\\
    W_{0,2}(x_1(z_1),x_2(z_2))=&\frac{1}{x_1'(z_1)x'_2(z_2)(z_1-z_2)^2}.
\end{align*}
\end{example}

\subsection{$x-y$ Symplectic Transformation}\label{Sec:x-y}
Symplectic transformations play a very important role in the theory of TR. All transformations which leaves the symplectic form 
\begin{align*}
    |dx\wedge dy|
\end{align*}
invariant are generated by the three transformations:
\begin{itemize}
    \item $(x,y)\to (x,y+R(x))$, where $R(x)$ is a rational function in $x$
    \item $(x,y)\to (\frac{ax+b}{cx+d},\frac{(cx+d)^2}{ad-bc}y)$ with $\begin{pmatrix}
		a & b \\
		c & d 
		\end{pmatrix}\in SL_2(\mathbb{C})$
	\item $(x,y)\to (y,x)$.
\end{itemize}
It was proved in \cite{Eynard:2007nq,Eynard:2013csa} that for meromorphic $x,y$ the \textbf{free energies are invariant (up to a known normalisation constant) under all symplectic transformations}. 

The same is \textbf{not true for $\omega_{g,n}$}! Note however that the first two symplectic transformations listed above leave indeed $\omega_{g,n}$ for $\chi<0$ invariant. This happens due to the fact that $x$ and $y$ enter the recursion in \eqref{eq:TR-intro} just through the recursion kernel and the ramification points $\beta_i$ of $x$ which are both invariant. New inside was achieved recently on the third symplectic transformation, the so-called \textit{$x-y$ symplectic transformation}.  

In a series of papers \cite{Bychkov:2020yzy,Bychkov:2022wgw,JEP_2022__9__1121_0}, Bychkov et al. derived an involved formula for two different sets of connected moments coming from topological partition functions. This functional relation between two families of moments was proved to be the $x-y$ symplectic transformation for a certain type of matrix models \cite{Bychkov:2021hfh}. Due to the relation between random matrix theory and free probability, this functional relation was shown \cite{Borot:2021thu} to give the moment-cumulant relation in free probability. A fairly simple version of the same functional relation for genus $g=0$ was derived in \cite{Hock:2022wer} with a completely different technique via loop insertion operator, and for all genus in \cite{Hock:2022pbw}. Shortly later, Alexandrov et al. \cite{Alexandrov:2022ydc} have shown that the simplified version from 
\cite{Hock:2022pbw} is indeed the symplectic transformation for any meromorphic $x,y$.
 
To formulate the functional relation, we define for the spectral curve $(\Sigma,y,x,B)$ (now with $x$ and $y$ interchanged) the corresponding multi-differential with $\omega_{g,n}^\vee$. More precisely, $\omega_{0,1}^\vee(z)=x(z)dy(z)$, $\omega_{0,2}^\vee(z_1,z_2)=B(z_1,z_2)$ and all $\omega_{g,n}^\vee$ via \eqref{eq:TR-intro} with interchanged role of $x$ and $y$. This means $\omega_{g,n}^\vee$ has poles just located at the ramification points of $y$ for $\chi<0$. Similarly, we define in this setting
\begin{align}
    W^\vee_{g,n}(y_1(z_1),...,y_n(z_n))dy_1(z_1)...dy_n(z_n):=&\omega_{g,n}^\vee(z_1,...,z_n)\\
    W_n^\vee(y_1,...,y_n):=&\sum_{g=0}^\infty\hbar^{2g+n-2}W^\vee_{g,n}(y_1,...,y_n)\\
    \Phi^\vee_{g,n}(y_1(z_1),...,y_n(z_n)):=&\int_o^{z_1}...\int_o^{z_n}\omega_{g,n}^\vee(z_1,...,z_n)\\
    \Phi_n^\vee(y_1,...,y_n):=&\sum_{g=0}^\infty \hbar^{2g+n-2}\Phi_{g,n}^\vee(y_1,...,y_n).
\end{align}

Next, we need the following graphs describing the structure of the functional relation:
\begin{definition}\label{def:graph}
	Let $\mathcal{G}_{n}$ be the set of connected bicoloured graph $\Gamma$  with $n$ $\bigcirc$-vertices and $\bullet$-vertices, such that the following holds:
	\begin{itemize}
		\item[-] the  $\bigcirc$-vertices are labelled from $1,...,n$
		\item[-] edges are only connecting $\bullet$-vertices with $\bigcirc$-vertices
		\item[-] $\bullet$-vertices have valence $\geq 2$.
	\end{itemize}
	For a graph $\Gamma\in \mathcal{G}_{n}$, let $r_{i}(\Gamma)$ be the valence of the $i^{\text{th}}$ $\bigcirc$-vertex. 
	
	Let $I\subset \{1,...,n\}$ be the set associated to a $\bullet$-vertex, where $I$ is the set labellings of $\bigcirc$-vertices connected to this $\bullet$-vertex. Let $\mathcal{I}(\Gamma)$ be the set of all sets $I$ for a given graph $\Gamma\in \mathcal{G}_{n}$.
\end{definition}\noindent
The automorphism group $\mathrm{Aut}(\Gamma)$ consists of permutations of edges which preserve the structure of $\Gamma$ considering the labellings. A graph $\Gamma\in \mathcal{G}_n$ is up to automorphisms completely characterised by the set $\mathcal{I}(\Gamma)$.

Adapted to the definition and functions above, the functional relation reads in a very compact form:
\begin{theorem}[\cite{Hock:2022pbw,Alexandrov:2022ydc}]\label{Thm:FR}
Let $x,y$ be two meromorphic functions on a compact Riemann surface with simple distinct ramification points, which generates via TR \eqref{eq:TR-intro} the multi-differentials $\omega_{g,n}$ and $\omega_{g,n}^\vee$ as above. Let $\Phi_n,\Phi_n^\vee, W_n,W_n^\vee$ be as above defined from $\omega_{g,n}$ and $\omega_{g,n}^\vee$. For $I=\{i_1,...,i_n\}$, let 
\begin{align}\label{PhiHut}
	\hat{\Phi}^\vee_n(y_I;\hbar,u_I):=&\!\!\!\!\!\!\!\!\!\!\!\!\sum_{(\varepsilon_{i_1},...,\varepsilon_{i_n})\in\{ 1,-1\}^n}\!\!\!\!\!\!\!\!\!\!\!\! (-1)^{\#(\varepsilon_i=-1)} \Phi^\vee_n\bigg(y_{i_1}+\varepsilon_{i_1}\frac{\hbar u_{i_1}}{2},...,y_{i_n}+\varepsilon_{i_n}\frac{\hbar u_{i_n}}{2}\bigg)
\end{align}
and for $I=\{i,i\}$ (and genus zero spectral curve) the special case
\begin{align}\label{PhiHutDia}
    \hat{\Phi}^\vee_n(y_I;\hbar,u_I):=&\!\!\!\!\!\!\!\!\!\!\!\!\sum_{(\varepsilon_{1},\varepsilon_{2})\in\{ 1,-1\}^2}\!\!\!\!\!\!\!\!\!\!\!\! (-1)^{\#(\varepsilon_i=-1)} \bigg[\Phi^\vee_2\bigg(y_{i}+\varepsilon_{1}\frac{\hbar u_{i}}{2},y_{j}+\varepsilon_{2}\frac{\hbar u_{i}}{2}\bigg)\\\nonumber
    &\qquad \qquad\qquad -\log \bigg(y_{i}+\varepsilon_{1}\frac{\hbar u_{i}}{2}-y_{j}-\varepsilon_{2}\frac{\hbar u_{i}}{2}\bigg)\bigg]_{j=i}
\end{align}
(for higher genus spectral curves the logarithm has to be replaces by the appropriate Theta-function).
Let further be $\hat{O} $  a differential operator acting from the left
	\begin{align}\label{OOperator}
	\hat{O} (y_i):=&\sum_{m} \bigg(-\frac{\partial}{\partial {x_i}}\bigg)^m\bigg(-\frac{dy_i}{dx_i}\bigg)
	[u_i^m]\frac{\exp\bigg(\hat{\Phi}^\vee_1(y_i;\hbar,u_i)-x_iu_i\bigg)}{\hbar u_i}.
	\end{align}
    Then, the functional relation holds as a formal expansion in $\hbar$
    \begin{align*}\boxed{
        W_n(x_1(z_1),...,x_n(z_n))=\sum_{\Gamma\in\G_n}\frac{1}{|\mathrm{Aut}(\Gamma)|}\prod_{i=1}^n\hat{O} (y_i(z_i))\prod_{I\in\I(\Gamma)}\hat{\Phi}^\vee_n(y_I(z_I);\hbar,u_I).}
    \end{align*}
    \begin{proof}
        The functional relation stated in the theorem is slightly different from the one in \cite{Hock:2022pbw,Alexandrov:2022ydc}. First of all, we have changed the role of $x$ and $y$ comparing to \cite{Hock:2022pbw}. Next, as a formal expansion in $\hbar$, the weight function $\hat{\Phi}^\vee_n(y_I;\hbar,u_I)$ can be written as
        \begin{align*}
            \hat{\Phi}^\vee_n(y_I;\hbar,u_I)=&\!\!\!\!\!\!\!\!\!\!\!\!\sum_{(\varepsilon_{i_1},...,\varepsilon_{i_n})\in\{ 1,-1\}^n}\!\!\!\!\!\!\!\!\!\!\!\! (-1)^{\#(\varepsilon_i=-1)} \Phi^\vee_n\bigg(y_{i_1}+\varepsilon_{i_1}\frac{\hbar u_{i_1}}{2},...,y_{i_n}+\varepsilon_{i_n}\frac{\hbar u_{i_n}}{2}\bigg)\\
            =&\bigg(\prod_{i\in I}\hbar u_i S(\hbar u_i\partial_{x_i})\bigg)\big( W_{n}(x_I)\bigg),
        \end{align*}
        where $S(u)=\frac{e^{u/2}-e^{-u/2}}{u}$, see \cite{Hock:2022pbw} for more details. Expanding the lhs and the rhs in $\hbar$ gives for each coefficient the relation stated in \cite{Hock:2022pbw,Alexandrov:2022ydc}.
    \end{proof}
\end{theorem}
Now, we still want to understand more properties of the functional relation of Theorem \ref{Thm:FR}. 
\begin{example}\label{Ex:2}
Consider Example \ref{Ex:1} with $x,y$ interchanged, i.e. $y$ is unramified. All $W_{g,n}^\vee=0$ for $2g+n-2>0$. Therefore, all $\hat{\Phi}^\vee_n(y_I(z_I);\hbar,u_I)=0$ for $n>2$. Let $\mathcal{G}_n^2\subset \mathcal{G}_n$ be the set of graphs defined in Definition \ref{def:graph} with just 2-valent $\bullet$-vertices, then
\begin{align*}
        W_n(x_1(z_1),...,x_n(z_n))=\sum_{\Gamma\in\G_n^2}\frac{1}{|\mathrm{Aut}(\Gamma)|}\prod_{i=1}^n\hat{O} (y_i(z_i))\prod_{I\in\I(\Gamma)}\hat{\Phi}^\vee_2(y_I(z_I);\hbar,u_I).
    \end{align*}
Note that all diagonal $\hat{\Phi}^\vee_2(y_i(z_i),y_i(z_i);\hbar,u_i,u_i)$ as defined in \eqref{PhiHutDia} can be included in the exponential of the Operator $\hat{O}$ defined in \eqref{OOperator}. Also the other $\hat{\Phi}^\vee_2(y_i(z_i),y_j(z_j);\hbar,u_i,u_j)$ with $i\neq j$ can be collected as an exponential such that multiple $\bullet$-vertices connecting the same $\bigcirc$-vertices are generated through expansion of this exponential (this was already discussed in \cite[\S 7]{Alexandrov:2022ydc}). The symmetry factor becomes redundant. Thus, we have the alternative form
\begin{align*}
        W_n(x_1(z_1),...,x_n(z_n))=\prod_{i=1}^n\hat{O}^2 (y_i(z_i))\sum_{\Gamma\in\tilde{\G}_n^2}\prod_{I\in\I(\Gamma)}\bigg(e^{\hat{\Phi}^\vee_2(y_I(z_I);\hbar,u_I)}-1\bigg),
    \end{align*}
where $\tilde{\G}_n^2\subset \G_n^2\subset \mathcal{G}_n$ is the set of graphs defined in Definition \ref{def:graph} with 2-valent $\bullet$-vertices just connecting two different $\bigcirc$-vertices and at most one $\bullet$-vertex connects the same $\bigcirc$-vertices. Equivalently, $\tilde{\G}_n^2$ is the set of connected labeled graphs with $n$ vertices (\href{https://oeis.org/A001187}{A001187}). The modified operator is
\begin{align*}
    \hat{O}^2 (y_i)
    :=&\sum_{m } \bigg(-\frac{\partial}{\partial {x_i}}\bigg)^m\bigg(-\frac{dy_i}{dx_i}\bigg)
	[u_i^m]\frac{\exp\bigg(\hat{\Phi}^\vee_1(y_i;\hbar,u_i)-x_iu_i+\frac{1}{2}\hat{\Phi}^\vee_2(y_i,y_i;\hbar,u_i,u_i)\bigg)}{\hbar u_i}.
\end{align*}
The symmetry factor $\frac{1}{2}$ inside the exponential comes from the automorphism $\mathrm{Aut}(\Gamma)$ swapping the two edge of a single $\bullet$-vertex connected to the same $\bigcirc$-vertex. The permutation of $k$ $\bullet$-vertices connected to the same $\bigcirc$-vertex is $k!$ and also an automorphism in $\mathrm{Aut}(\Gamma)$, which is collected in the expansion of the exponentials.
\end{example}

The example shows that terms coming from the graph expansion can be nicely collected in an exponential. This is not a surprise since the original derivation came indeed from \textit{not-necessarily-connected correlators} \cite{JEP_2022__9__1121_0}. Rephrasing these computational steps backwards, a even more compact formula can be provided for the not-necessarily-connected correlators $\overset{\circ}{W}_n$ defined by
\begin{align}\label{Wdis}
    \overset{\circ}{W}_n(x_I):=\sum_{\lambda \vdash I}\prod_{i=1}^{l(\lambda)}W_{|\lambda_i|}(x_{\lambda_i}),
\end{align}
where $\lambda \vdash I$ is a set partition of $I=\{1,...,n\}$, i.e. $\lambda=\{\lambda_1,...,\lambda_{l(\lambda)}\}$ of length $l(\lambda)$ and blocks $\lambda_i\subset I$.  Note that $\overset{\circ}{W}_n(x_I)$ has for all $n>1$ and at each order in $\hbar$ in general poles on the diagonal on the variables $z_i,z_j$. 
\begin{corollary}\label{cor:disW}
The not-necessarily-connected correlators $\overset{\circ}{W}_n$ satisfy the functional relation as formal expansion in $\hbar$
\begin{align}\label{disconW}
    \overset{\circ}{W}_n&(x_I(z_I))=\sum_{m_1,...,m_n}\prod_{i=1}^n\bigg(-\frac{\partial}{\partial {x_i(z_i)}}\bigg)^{m_i}\bigg(-\frac{dy_i(z_i)}{dx_i(z_i)}\bigg)
	[u_i^{m_i}]\frac{1}{\hbar u_i}\\\nonumber
 &\times\exp\bigg(\sum_{k\geq 1}\frac{1}{k!}\sum_{i_1,...,i_k=1}^n\hat{\Phi}^\vee_{k}(y_{i_1}(z_{i_1}),...,y_{i_k}(z_{i_k});\hbar,u_{i_1},...,u_{i_k})\\\nonumber
 &\qquad \qquad -\sum_{i=1}^n u_i x_i(z_i)\bigg).
\end{align}
\begin{proof}
    Since $\overset{\circ}{W}_n(x_I)$ includes all not-necessarily-connected correlators, it is a rather classical result that the exponential generates these from the connected correlators (very similar to the discussion in Example \ref{Ex:2}). The automorphisms are split in two groups. The first is generating the symmetry factors $\frac{1}{k_i!}$ permuting a $\bullet$-vertex connected with $k_i$ edges to the $i$-th $\bigcirc$-vertex. The second permutes the same $\bullet$-vertices with the same number of edge connecting to the some $\bigcirc$-vertices, which is collected in the expansion of the exponential. Since all $\hat{\Phi}^\vee_{k}$ are symmetric, we can reorder the summation in the exponential (via multinomial theorem) as
    \begin{align*}
        &\sum_{\substack{k_1,...,k_n\geq 0\\ k_1+...+k_n=k>0}}\frac{\hat{\Phi}^\vee_{k}(\overbrace{y_1(z_1),...,y_1(z_1)}^{k_1},...,\overbrace{y_n(z_n),...,y_n(z_n)}^{k_n} ;\hbar,u_1,...)}{k_1!...k_n!}\\
        =&\sum_{k\geq 1}\frac{1}{k!}\sum_{i_1,...,i_k=1}^n\hat{\Phi}^\vee_{k}(y_{i_1}(z_{i_1}),...,y_{i_k}(z_{i_k});\hbar,u_{i_1},...,u_{i_k}),
    \end{align*}
        and get the claimed result.
\end{proof}
\end{corollary}
We observe that the not-necessarily-connected correlator $\overset{\circ}{W}_n$, which are generated by the spectral curve $(x,y)$, are related to the connected correlators $W_{g,n}^\vee$ generated by the spectral curve $(y,x)$ through a differential operator acting on an exponential. However, the form of the differential operator taking as $m$-th derivative of the $m$-th order in the $u$ expansion can have a more transparent explanation. Furthermore, we will also give insight on the factor $\bigg(-\frac{dy_i(z_i)}{dx_i(z_i)}\bigg)$ through some formal observations.

\subsection{Laplace transform of the $x-y$ symplectic transformation}\label{sec:laplace}
Let us look at the \textit{Laplace transform} of \eqref{disconW}. The integration path $\gamma$ on the Riemann surface depends on $x(z),y(z)$. We will not be precise about this path but want rather look at the following formal manipulations:
\begin{proposition}\label{Pro:Laplace}
    Assume paths $\gamma_i\ni z_i$ exist such that $\overset{\circ}{W}_n(x_1(z_1),...,x_n(z_n))$ is analytic on $\gamma_i$ and the integrand vanishes fast enough at its boundary values. Assume further that  the Laplace transform of $\overset{\circ}{W}_n(x_I)$ along the paths $\gamma_i$ converges for each coefficient in $\hbar$, then the Laplace transform as a formal expansion in $\hbar$ reads
    \begin{align*}
        &\int_{\gamma_1} dx_1(z_1)e^{-\mu_1x_1(z_1)}...\int_{\gamma_n}dx_n(z_n)e^{-\mu_nx_n(z_n)} \overset{\circ}{W}_n(x_1(z_1),...,x_n(z_n))\\
        =&\int_{\gamma_1} \frac{dy_1(z_1)}{\hbar \mu_1}...\int_{\gamma_n}\frac{dy_n(z_n)}{\hbar \mu_n}\exp\bigg(\sum_{k\geq 1}\frac{1}{k!}\sum_{i_1,...,i_k=1}^n\hat{\Phi}^\vee_{k}(y_{i_1}(z_{i_1}),...,y_{i_k}(z_{i_k});\hbar,-\mu_{i_1},...,-\mu_{i_k})\bigg).
    \end{align*}
    \begin{proof}
        Take Corollary \ref{cor:disW} and multiply it with $\prod_{i=1}^ne^{-\mu_ix_i(z_i)}dx_i(z_i)$ and integrate over $\gamma_i$ in each variable $z_i$. Due to the assumptions, we can integrate by parts in each variable $x_i(z_i)$ exactly $m_i$ times such that all boundary terms vanish. This yields
        \begin{align*}
        &\int_{\gamma_1} dx_1(z_1)e^{-\mu_1x_1(z_1)}...\int_{\gamma_n}dx_n(z_n)e^{-\mu_nx_n(z_n)} \overset{\circ}{W}_n(x_1(z_1),...,x_n(z_n))\\
            =&\int_{\gamma_1} dx_1(z_1)e^{-\mu_1x_1(z_1)}...\int_{\gamma_n}dx_n(z_n)e^{-\mu_nx_n(z_n)}\sum_{m_1,...,m_n\geq 0}\prod_{i=1}^n\bigg(-\mu_i\bigg)^{m_i}\bigg(-\frac{dy_i(z_i)}{dx_i(z_i)}\bigg)
	[u_i^{m_i}]\\\nonumber
 &\times\frac{1}{\hbar u_i}\exp\bigg(\sum_{k\geq 1}\frac{1}{k!}\sum_{i_1,...,i_k=1}^n\hat{\Phi}^\vee_{k}(y_{i_1}(z_{i_1}),...,y_{i_k}(z_{i_k});\hbar,u_{i_1},...,u_{i_k}) -\sum_{i=1}^n u_i x_i(z_i)\bigg).
        \end{align*}
        As a formal expression in $u_i$, we can just substitute all $u_i$ with $-\mu_i$ and ignore the $m_i$ summation. The factors $e^{-\mu_1x_1(z_1)}$ are cancelled by $e^{-\sum_{i=1}^n u_i x_i(z_i)}$ for $u_i=-\mu_i$. Changing the integration variable $dx_i(z_i)\to dy_i(z_i)$ finishes the proof.
    \end{proof}
\end{proposition}
The proposition gives the most compact and transparent relation between the correlators $W_{g,n}$ and $W_{g,n}^\vee$ generated via TR with spectral curve $(x,y)$ and $(y,x)$, respectively. We see explicitly that taking the $m_i$-th derivative of a formal expansion of the $m_i$-th coefficient is a very natural operation for it Laplace transform. The appearance of the factor $-\frac{dy_i(z_i)}{dx_i(z_i)}$ yields the final change of integration variables. However, the Laplace transform depends on the different paths $\gamma_i$, and interchanging them  $\gamma_i\to \gamma_{\sigma(i)}$ under some permutation gives first of all a different Laplace transform, since $\overset{\circ}{W}_n$ has poles on the diagonal and residues at the diagonal will not necessarily vanish.

\begin{example}\label{Ex:leadingOrder}
The leading order in $\hbar$ of Proposition \ref{Pro:Laplace} is $\hbar^{-n}$. The lhs of the equation expands into $\overset{\circ}{W}_n(x_1(z_1),...,x_n(z_n))=\prod_{i=1}^n\frac{y_i(z_i)}{\hbar}+\mathcal{O}(\hbar^{-n+1})$, and for the rhs the argument of the exponential expands at leading order to 
\begin{align*}
    &\sum_{i=1}^n\hat{\Phi}^\vee_{1}(y_i(z_i);\hbar;-\mu_i)+\mathcal{O}(\hbar)\\
    =&\frac{1}{\hbar}\sum_{i=1}^n \bigg(\Phi^\vee_{1}\bigg(y_i(z_i)-\frac{\hbar \mu_i}{2}\bigg)-\Phi^\vee_{1}\bigg(y_i(z_i)+\frac{\hbar \mu_i}{2}\bigg)\bigg)+\mathcal{O}(\hbar)\\
    =&-\sum_{i=1}^n\mu_i x_i(z_i)+\mathcal{O}(\hbar).
\end{align*}
Putting everything together, the leading order reads
\begin{align*}
    &\int_{\gamma_1} dx_1(z_1)e^{-\mu_1x_1(z_1)} y_1(z_1)...\int_{\gamma_n}dx_n(z_n)e^{-\mu_nx_n(z_n)}y_n(z_n)\\
    =&\int_{\gamma_1} \frac{dy_1(z_1)}{\mu_1}e^{-\mu_1x_1(z_1)} ...\int_{\gamma_n}\frac{dy_n(z_n)}{\mu_n}e^{-\mu_nx_n(z_n)}
\end{align*}
which is correct under the assumptions and integration by parts.
\end{example}

\begin{example}\label{Ex:3}
Consider the same situation as in Example \ref{Ex:1} and \ref{Ex:2}, this is $y$ is unramified thus all $W_{g,n}^\vee=0$ for $2g+n-2>0$. As before, this implies that all $\hat{\Phi}^\vee_{k}=0$ for $k>2$. We conclude for this case from Proposition \ref{Pro:Laplace}
\begin{align*}
    &\int_{\gamma_1} dx_1(z_1)e^{-\mu_1x_1(z_1)}...\int_{\gamma_n}dx_n(z_n)e^{-\mu_nx_n(z_n)} \overset{\circ}{W}_n(x_1(z_1),...,x_n(z_n))\\
        =&\int_{\gamma_1} \frac{dy_1(z_1)}{\hbar \mu_1}...\int_{\gamma_n}\frac{dy_n(z_n)}{\hbar \mu_n}\exp\bigg(\sum_{i=1}^n\hat{\Phi}^\vee_{1}(y_i(z_i);\hbar;-\mu_i)+\frac{1}{2}\sum_{i,j=1}^n \hat{\Phi}^\vee_{2}(y_i(z_i),y_j(z_j);\hbar;-\mu_i,-\mu_j)
        \bigg).
\end{align*}
Note that on the diagonal $\hat{\Phi}^\vee_{2}(y_i(z_i),y_i(z_i);\hbar;\mu_i,\mu_i)$, we have to take the normalised primitives \eqref{PhiHutDia}.
\end{example}

The derivation above is also valid for connected correlators $W_n$ with exactly the same steps. We just state the result:
\begin{corollary}\label{cor:LaplaceConW}
    Assume a paths $\gamma_i$ exist such that $W_{n}(x_1(z_1),...,x_n(z_n))$ is analytic on $\gamma_i$ and the integrand vanishes fast enough at its boundary values. Assume the Laplace transform of ${W}_n(x_I)$ along the path $\gamma$ converges for each coefficient in $\hbar$, then the Laplace transform as a formal expansion in $\hbar$ reads
    \begin{align*}
        &\int_{\gamma_1} dx_1(z_1)e^{-\mu_1x_1(z_1)}...\int_{\gamma_n}dx_n(z_n)e^{-\mu_nx_n(z_n)} {W}_n(x_1(z_1),...,x_n(z_n))\\
        =&\int_{\gamma_1} \frac{dy_1(z_1)}{\hbar \mu_1}...\int_{\gamma_n}\frac{dy_n(z_n)}{\hbar \mu_n}\exp\bigg(\sum_{i=1}^n\hat{\Phi}^\vee_{1}(y_{i}(z_{i});\hbar,-\mu_{i})\bigg)\sum_{\Gamma\in\G_n}\frac{\prod_{I\in\I(\Gamma)}\hat{\Phi}^\vee_{|I|}(y_I(z_I);\hbar,-\mu_I)}{|\mathrm{Aut}(\Gamma)|}.
    \end{align*}
\end{corollary}

\begin{example}\label{Ex:4}
    Continue the series of Example \ref{Ex:1}, \ref{Ex:2} and \ref{Ex:3}, where $y$ is unramified. The Laplace transform of the connected correlator reads
    \begin{align*}
        &\int_{\gamma_1} dx_1(z_1)e^{-\mu_1x_1(z_1)}...\int_{\gamma_n}dx_n(z_n)e^{-\mu_nx_n(z_n)} {W}_n(x_1(z_1),...,x_n(z_n))\\
        =&\int_{\gamma_1} \frac{dy_1(z_1)}{\hbar \mu_1}...\int_{\gamma_n}\frac{dy_n(z_n)}{\hbar \mu_n}\exp\bigg(\sum_{i=1}^n\hat{\Phi}^\vee_{1}(y_{i}(z_{i});\hbar,-\mu_{i})+\frac{1}{2}\hat{\Phi}^\vee_{2}(y_{i}(z_{i}),y_{i}(z_{i});\hbar,-\mu_{i},-\mu_i)\bigg)\\
        &\qquad \qquad \qquad \times\sum_{\Gamma\in\tilde{\G}^2_n}\prod_{I\in\I(\Gamma)}\bigg(e^{\hat{\Phi}^\vee_2(y_I(z_I);\hbar,-\mu_I)}-1\bigg),
    \end{align*}
    where $\tilde{\G}^2_n\subset \G_n$ is the set of graphs with 2-valent $\bullet$-vertices connecting two different $\bigcirc$-vertices at most with one $\bullet$-vertex which is nothing than the set of connected graphs with $n$ labelled vertices (\href{https://oeis.org/A001187}{A001187}).
\end{example}

\begin{remark}
    The definition of the Laplace transform in several variables indicates a possible dependence on the different paths $\gamma_i$. Permuting or geometrically moving two paths $\gamma_i,\gamma_j$ along each other would pick a non-trivial residue at the  diagonal $z_i=z_j$. However, it turns out that after computing all path integrals these residues cancel out. This is related to the fact that the lhs of the functional relation of Theorem \ref{Thm:FR} has no pole at the diagonal, even though the rhs is indicating it.
\end{remark}

\subsection{Logarithmic $x,y$}\label{sec:logxy}
In the previous subsection, logarithmic behaviour of $x$ or $y$ was excluded. However, the TR itself can also be applied to logarithmic $x$ and $y$. Actually, a lot of very important examples related for instance to Gromov-Witten theory have logarithms. To verify that the functional relation does not hold in general for logarithmic $x,y$, we take the example of the so-called \textit{Lambert curve} \cite{Bouchard:2007hi,Eynard2009TheLT}. This curve encodes Hurwitz numbers, which will be explained in more details later. The curve is defined via 
\begin{align}\label{lam1}
    x(z)= -z+\log(z),\qquad y(z)=z.
\end{align}
The spectral curve is of genus zero such that the bilinear differential is $\omega_{0,2}(z_1,z_2)=\frac{dz_1\,dz_2}{(z_1-z_2)^2}$. Starting with these two functions, the formula of TR  \eqref{eq:TR-intro} generates all $W_{g,n}$ and also $W^\vee_{g,n}$.  One might check if the functional relation holds for some examples.

Since $y$ is unramified, this curve reflects the examples of the previous subsection, i.e. $W^\vee_{g,n}=0$ for all $-\chi<2g+n-2$. 
A short computation of \eqref{eq:TR-intro} gives for instance 
\begin{align*}
    W_{1,1}(x(z))=\frac{1}{x'(z)}
  \Res\displaylimits_{q\to 1}
  K_i(z,q)
  \omega_{0,2}( q,\sigma_i(q))=\frac{z^2(z-4)}{24 (z-1)^5}.
\end{align*}
On the other hand, the $(g,n)=(1,1)$-example of the functional relation is 
\begin{align*}
	W_{1,1}(x(z))=&-\frac{dy(z)}{dx(z)}W^\vee_{1,1}(y(z))+\frac{1}{2}\frac{d}{dx(z)}\bigg(\frac{dy(z)}{dx(z)}\hat{W}^\vee_{2,0}(y(z),y(z))\bigg)
	-\frac{1}{24}\frac{d^3}{dx(z)^3}\bigg(\frac{1}{\frac{dy(z)}{dx(z)}}\bigg)\\
 =&-\frac{1}{24}\frac{d^3}{dx(z)^3}\bigg(\frac{1}{\frac{dy(z)}{dx(z)}}\bigg)
 \\
 =&\frac{-6z^2+4z-1}{24 z(z-1)^5}
\end{align*}
where $W^\vee_{1,1}(y(z))=0$ and $\hat{W}^\vee_{2,0}(y_1,y_2)=W^\vee_{2,0}(y_1,y_2)-\frac{1}{(y_1-y_2)^2}=0$ vanish such that the last term contributes only. This is a clear discrepancy to the direct computation.

However, if we look at a symplectic transformation of this curve by transforming 
\begin{align}\label{symlam}
    y\to \tilde{y}=y+x,
\end{align}
the validity of the functional relation can be rescued. This means we start with the curve
\begin{align}\label{lam2}
    x(z)= -z+\log(z),\qquad \tilde{y}(z)=\log(z).
\end{align}
Let us denote the corresponding correlators with $\tilde{W}_{g,n}$, which are equal to $W_{g,n}$ of the curve \eqref{lam1} due to symplectic transformation. Thus, the formula of TR yields obviously again
\begin{align*}
    \tilde{W}_{1,1}(x(z))=\frac{z^2(z-4)}{24 (z-1)^5}
\end{align*}
due to the invariance of the kernel under this transformation. After $x-y$ symplectic transformation, i.e. looking at the correlators $\tilde{W}^\vee_{g,n}$, we see that $\tilde{W}^\vee_{g,n}=0$ for all $-\chi<2g+n-2$ since $y$ is unramified again. However, the regularised correlators $\hat{\tilde{W}}^\vee_{0,2}$ on the diagonal does not vanish. It is
\begin{align*}
    \lim_{z'\to z}\frac{1}{\tilde{y}'(z)\tilde{y}'(z')(z-z')^2}-\frac{1}{(\tilde{y}(z)-\tilde{y}(z'))^2}=-\frac{1}{12}.
\end{align*}
Inserting everything into the functional relation for $(g,n)=(1,1)$, we find
\begin{align*}
	\tilde{W}_{1,1}(x(z))=&-\frac{d\tilde{y}(z)}{dx(z)}\tilde{W}^\vee_{1,1}(y(z))+\frac{1}{2}\frac{d}{dx(z)}\bigg(\frac{d\tilde{y}(z)}{dx(z)}\hat{\tilde{W}}^\vee_{2,0}(\tilde{y}(z),\tilde{y}(z))\bigg)
	-\frac{1}{24}\frac{d^3}{dx(z)^3}\bigg(\frac{1}{\frac{d\tilde{y}(z)}{dx(z)}}\bigg)\\
 =&-\frac{1}{24}\frac{d}{dx(z)}\bigg(\frac{d\tilde{y}(z)}{dx(z)}\bigg)-\frac{1}{24}\frac{d^3}{dx(z)^3}\bigg(\frac{1}{\frac{d\tilde{y}(z)}{dx(z)}}\bigg)
 \\
 =&\frac{z^2(z-4)}{24 (z-1)^5}
\end{align*}
which coincides with the direct computation from TR.

This example reveals that considering logarithmic $x,y$ the $x-y$ symplectic transformation does not hold in general. But on the other hand, taking the correct symplectic transformation \eqref{symlam} before, it actually can hold. One may ask if the Laplace transform of Proposition \ref{Pro:Laplace} and Corollary \eqref{cor:LaplaceConW} is still valid if the functional relation is satisfied including logarithms for $x,y$ as it is for the curve \eqref{lam2}.

\begin{remark}\label{rem:01}
For logarithmic behaviour of $x,y$, the contour $\gamma$ may cross the branch cut of the logarithm. The observation of Example \ref{Ex:leadingOrder} gives us an alternative way making sense of the Laplace transform even for $(g,n)=(0,1)$ including logarithms for $x,y$. Therefore, we define 
\begin{align*}
    \int_\gamma dx(z)e^{-\mu x(z)} y(z)
    :=\int_\gamma \frac{dy(z)}{\mu}e^{-\mu x(z)} ,
\end{align*}
if $y$ has a logarithm and $\gamma$ crosses the branch cut. This avoids to split the integration contour at the cut and include boundary terms to regularise the lhs, which is well-defined and could be done in principle.
\end{remark}

\section{Application to Intersection Numbers on $\overline{\mathcal{M}}_{g,n}$}
This section recalls some examples about the connection between TR and intersection theory on $\overline{\mathcal{M}}_{g,n}$ and applies the derived formulas. We refer to \cite{Eynard:2011kk,Eynard2011InvariantsOS} for much more information.  

Let $\overline{\mathcal{M}}_{g,n}$ be the compactified moduli space of complex curves of genus $g$ with $n$ labelled points. It is compactified in the sense Deligne and Mumford \cite{MR262240}. $\overline{\mathcal{M}}_{g,n}$ is a complex orbifold of dimension $d_{g,n}=3g-3+n$. Its points $(C,p_1,...,p_n)\in \overline{\mathcal{M}}_{g,n}$ are isomorphism classes of a complex curve $C$ with $n$ labelled point denoted by $p_1,...,p_n$. Let $L_i$ be the line bundle over $\overline{\mathcal{M}}_{g,n}$ whose fiber is the contangent space $TC^\vee(p_i)$ of $C$ at $p_i$. The first Chern class of this line bundle is called the $\psi$-class
\begin{align*}
    \psi_i=c_1(L_i).
\end{align*}
The $\psi$-classes are the easiest examples for tautological classes in $\overline{\mathcal{M}}_{g,n}$. One can build \textit{intersection numbers} by wedging several $\psi$-classes
\begin{align}\label{PsiIntersec}
    \langle\psi_1^{d_1}...\psi_n^{d_n}\rangle_{g,n}=\int_{\overline{\mathcal{M}}_{g,n}}\psi_1^{d_1}...\psi_n^{d_n}
\end{align}
with $d_{g,n}=\sum_{i=1}^nd_i$, otherwise it is defined to vanish. It was conjectured by Witten \cite{Witten:1990hr} and shortly later proved by Kontsevich \cite{Kontsevich:1992ti} that the generating function \eqref{PsiIntersec} satisfies the KdV equations, see for instance \cite{Lando,Eynard:2016yaa} for more details.

Kontsevich proved that a hermitian matrix model with external field generates the intersection numbers \eqref{PsiIntersec} if one identify the parameters of their generating function with certain coefficients of the matrix model. After developing TR, this connection constructed by Kontsevich between matrix models and intersection theory gave  birth to the connection between TR and intersection theory in general. The associated spectral curve $\psi$-class intersection numbers is known under the name of Airy curve or Witten-Kontsevich curve, defined by $(\mathbb{P}^1,,x=\frac{z^2}{2},y=z,\frac{dz_1\,dz_2}{(z_1-z_2)^2})$ \cite{Eynard:2007kz}.

We will also discuss more general intersection numbers built from $\psi$- and Hodge-classes. Let $\pi:\overline{\mathcal{M}}_{g,n+1}\to \overline{\mathcal{M}}_{g,n}$ be the forgetful morphism, forgetting the last labelled point, and $\omega_{\pi}$ the relative dualising sheaf. Then, the Hodge-class is defined by
\begin{align}
    \Lambda(\alpha)=1+\sum_{k=1}^g(-1)^k\alpha^{-k}c_k(\mathbb{E}),
\end{align}
where $c_k$ is the $k$-th Chern class and $\mathbb{E}$ the Hodge bundle given by pushforward of the relative dualising sheaf, i.e. $\mathbb{E}=\pi_*(\omega_\pi)$.

Intersection numbers of a mixture of $\psi$- and Hodge-classes can be considered and made considerable interest in the past. Especially in the context of Hurwitz numbers, a relation between intersection numbers of $\psi$- and Hodge-classes and the counting problem of ramified coverings over $\mathbb{P}^1$ with a certain ramification profil at infinity. This relation is the celebrated ELSV formula \cite{EKEDAHL19991175}. More precisely, let $h_{g;k\mu_1,...,\mu_n}$ be the 
number of the equivalence classes of topologically nonequivalent ramified
coverings $f:C\to\mathbb{P}^1$, where $C$ is compact, connected complex curve of genus $g$ and $f$ has ramification profil $(\mu_1,....,\mu_n)$ over infinity and simple ramification else. The ELSV formula relates the Hurwitz number $h_{g;\mu_1,...,\mu_n}$ to the following linear Hodge integral
\begin{align}
    h_{g;\mu_1,...,\mu_n}=\frac{(2g+\mu+n-2)!}{|\mathrm{Aut}(\mu_1,...,\mu_n)|}\prod_{i=1}^n\frac{\mu_i^{\mu_i}}{\mu_i!}\int_{\overline{\mathcal{M}}_{g,n}}\frac{\Lambda(1)}{\prod_{i=1}^n(1-\mu_i\psi_i)},
\end{align}
where $|\mathrm{Aut}(\mu_1,...,\mu_n)|$ is the number of permutations permuting equal $\mu_i$'s and $\mu=\mu_1+...+\mu_n$. 

Bouchard and Marino conjectured \cite{Bouchard:2007hi} that these linear Hodge integrals appearing in the ELSV formula  (and therefore also Hurwitz numbers) can actually be computed by TR with the so-called \textit{Lambert curve}. This was proved in \cite{Eynard2009TheLT}, where the spectral curve was given by $(\mathbb{P}^1\setminus \mathbb{R}_-,x=-z+\log(z),y=\log(z),\frac{dz_1\,dz_2}{(z_1-z_2)^2})$. Note that we have shifted $y$ in the spectral curve due to the observation in Sec. \ref{sec:logxy}, which does not change the result of \cite{Eynard2009TheLT}.

Applying the Laplace transform of Corollary \ref{cor:LaplaceConW} to the Airy and/or Lambert curve gives easy formulas to compute $\psi$- and Hodge-class intersection numbers.

\subsection{Airy curve}
For the Airy curve, the explicit relation between the correlators $\omega_{g,n}$ and the intersection numbers is given by:
\begin{theorem}[\cite{Eynard:2007kz}]\label{Thm:Airy}
    Let the Airy spectral curve be $(\mathbb{P}^1,x(z)=z^2,y(z)=z, \frac{dz_1\,dz_2}{(z_1-z_2)^2})$, then the correlator $\omega_{g,n}$ computed by TR  generates the $\psi$-class integral
    \begin{align}\label{AiryCurve}
        \omega_{g,n}(z_1,...,z_n)=\sum_{k_1+...+k_n=d_{g,n}}\langle\psi_1^{k_1}...\psi_n^{k_n}\rangle_{g,n} \prod_{i=1}^n \frac{(2k_i+1)!!}{z_i^{2k_i+2}}dz_i.
    \end{align}
\end{theorem}
The first step is two extract intersection numbers from $\omega_{g,n}$ through a Laplace transform:
\begin{lemma}\label{lem:airy}
Let $\mu_i\in \mathbb{N}$ and $2g+n-2>0$, we have for $\omega_{g,n}$ generated by the Airy curve of Theorem \ref{Thm:Airy}
\begin{align}\label{eq:airy}
    \frac{1}{\sqrt{2\pi}^n}\int_{i \varepsilon-\infty}^{i \varepsilon+\infty}  e^{-\mu_1x_1(z_1)...-\mu_nx_n(z_n)} \omega_{g,n}(z_1,...,z_n)=\bigg\langle\prod_{i=1}^n\frac{\sqrt{\mu_i}}{1-\mu_i \psi_i}\bigg\rangle_{g,n}.
\end{align}
For $2g+n-2\leq 0$, we get $\bigg\langle\frac{1}{(1-\mu \psi)}\bigg\rangle_{0,1}=\frac{1}{\mu^2}$ and $\bigg\langle\frac{1}{(1-\mu_1\psi_1)(1-\mu_2\psi_2)}\bigg\rangle_{0,2}=\frac{1}{\mu_1+\mu_2}$.
\begin{proof}
    For $2g+n-2>0$ and each $\mu_i$, we apply the integral to the result of Theorem \ref{Thm:Airy}. The integral is computed by a version of the well-known Gaussian integral
 \begin{align}\label{GausInt}
        \frac{1}{\sqrt{2\pi }}\int_{i \varepsilon-\infty}^{i \varepsilon+\infty}dz \frac{e^{-\mu z^2/2}}{z^{2k+2}}=\frac{\mu^{k+1/2}}{(2k+1)!!}.
    \end{align}
    The sum over $k_1+...+k_n=d_{g,n}$ can be extended to a sum over all $k_i\geq 0$, since all the intersection number are defined to vanish unless the condition $k_1+...+k_n=d_{g,n}$ is satisfied. For each $k_i$ summation, we have a geometric series
    $\sum_{k_i=0}^\infty \mu_i^{k_i} \psi^{k_i}=\frac{1}{1-\mu_i \psi_i}$.
    
    For $2g+n-2\leq 0$, the computation is also straightforward. The Bergman kernel is first expanded in a geometric series $\frac{1}{(z_1-z_2)^2}=\frac{1}{z_1^2}\sum_n n (\frac{z_2}{z_1})^{n-1}$, then apply the Gaussian integral together with $\frac{1}{(2k-1)!! (-2k-1)!!}=(-1)^k$ and write it again as geometric series again.
\end{proof}
\end{lemma}

Now, we can apply the formulas developed in Sec. \ref{sec:laplace} to the Airy curve. First of all note that the Airy curve  $x(z)=z^2,y(z)=z$
is not ramified through $y$. We can easily stick to the examples discussed in Sec. \ref{sec:laplace} and just have to compute the following primitives:
\begin{align}\label{Phi1A}
    \Phi_1^\vee(y(z))&=\frac{1}{\hbar}\int x(z)dy(z)=\frac{z^3}{6 \hbar }\\\label{Phi1HA}
    \hat{\Phi}^\vee_1(y(z);\hbar,u)&=\Phi_1^\vee\bigg(y(z)+\frac{\hbar u}{2}\bigg)-\Phi_1^\vee\bigg(y(z)-\frac{\hbar u}{2}\bigg)=
    u\frac{z^2}{2}+\frac{\hbar^2u^3}{24}\\\label{Phi2A}
    \Phi_2^\vee(y_1(z_1),y_2(z_2))&=\log(z_1-z_2)\\\label{Phi1HA2}
    e^{\hat{\Phi}_2^\vee(y_1(z_1),y_2(z_2);\hbar,u_1,u_2)}-1&=\frac{\hbar^2 u_1 u_2}{(z_1-z_2)^2-\frac{\hbar^2}{4}(u_1+u_2)^2}.
\end{align}
Taking Example \ref{Ex:4} and Lemma \ref{lem:airy} into account, we conclude:
\begin{corollary}\label{cor:airy1}
    The $\psi$-class intersection numbers can be computed as a formal expansion in $\hbar$ via
    \begin{align*}
        &\bigg\langle\prod_{i=1}^n\frac{\sqrt{\mu_i}}{1-\mu_i \psi_i}\bigg\rangle_{g,n}\\
        =&[\hbar^{2g+n-2}] \prod_{i=1}^n\frac{e^{-\frac{\hbar^2\mu_i^3}{24}}}{\sqrt{2\pi}}\int_{i \varepsilon-\infty}^{i \varepsilon+\infty}dz_i \frac{e^{-\mu_i \frac{z_i^2}{2}}}{\hbar \mu_i}
        \sum_{\Gamma\in\tilde{\G}^2_n}\prod_{\substack{I\in\I(\Gamma)\\\{i,j\}=I}}\frac{\hbar^2 \mu_i \mu_j}{(z_i-z_j)^2-\frac{\hbar^2}{4}(\mu_i+\mu_j)^2},
    \end{align*}
    where $\tilde{\G}^2_n\subset \G_n$ is the subset of graphs with 2-valent $\bullet$-vertices connecting two different $\bigcirc$-vertices at most with one $\bullet$-vertex which is nothing than the set of connected graphs with $n$ labelled vertices (\href{https://oeis.org/A001187}{A001187}).
\end{corollary}
For not-necessarily-connected correlators, we can conclude from Example \ref{Ex:3} and Lemma \ref{lem:airy}:
\begin{corollary}\label{cor:airy2}
    The sum over all partitions of $\psi$-class intersection numbers satisfies as a formal expansion in $\hbar$
    \begin{align*}
        &\sum_{\lambda\vdash I}\prod_{j=1}^{l(\lambda)}\sum_{g_j=0}\hbar^{2g_j+|\lambda_j|-2} \bigg\langle\prod_{i=1}^{|\lambda_j|}\frac{\sqrt{\mu_{\lambda_j^i}}}{(1-\mu_{\lambda_j^i}\psi_{\lambda_j^i})}\bigg\rangle_{g_j,|\lambda_j|}\\
        =&\prod_{i=1}^n\frac{e^{-\frac{\hbar^2\mu_i^3}{24}}}{\sqrt{2\pi}}\int_{i \varepsilon-\infty}^{i \varepsilon+\infty}dz_i \frac{e^{-\mu_i \frac{z_i^2}{2}}}{\hbar \mu_i}
        \prod_{i<j}\frac{\hbar^2 \mu_i \mu_j}{(z_i-z_j)^2-\frac{\hbar^2}{4}(\mu_i+\mu_j)^2},
    \end{align*}
    where $\lambda\vdash I$ is a set partition of $I$ with $l(\lambda)$ blocks $\lambda_j\subset I$, i.e. $\lambda=(\lambda_1,...,\lambda_{l(\lambda)})$. Each block is written as $\lambda_j=(\lambda_j^1,...,\lambda_j^{|\lambda_j|})$ of cardinality $|\lambda_j|$ and elements $\lambda_j^i\in I$.
\end{corollary}

The result of Corollary \ref{cor:airy1} and \ref{cor:airy2} are not claimed to be new formulas, because these are equivalent to the one given in \cite[\S 7]{Alexandrov:2022ydc} after applying the Laplace transform as explained in Sec. \ref{sec:laplace}. So, performing the Gaussian integrals, one usually manipulate the integrand through derivatives to get formulas like \eqref{GausInt}, we just arrive at the formula directly induced by Theorem \ref{Thm:Airy} together with Theorem \ref{Thm:FR} as stated in \cite[\S 7]{Alexandrov:2022ydc}. However, we are still getting a new perspective on the computation of $\psi$-class intersection numbers in terms of Gaussian integrals. 

We want to emphasise also that the order of integration in Corollary \ref{cor:airy1} and \ref{cor:airy2} does not matter. This is related to the fact that the correlators $\omega_{g,n}$ have just poles at the ramification points and not at the diagonal for $2g+n-2>0$. The deep algebraic structure which reveals this property is not yet understood, but it is present in the background of all these formulas.

\subsection{Lambert curve}
In this subsection, we apply the formulas of Sec. \ref{sec:laplace} to the Lambert curve. Note that these formulas are just proved for meromorphic $x,y$. The discussion of Sec. \ref{sec:logxy} has however shown that for the Lambert curve a specific choice of $y$ can actually work. Through this subsection, we have assigned an asterisk to those corollaries which assume that for the Lambert curve of the form of \eqref{lam2} satisfies the $x-y$ symplectic transformation. A lot of checks with computer algebra have confirmed this assumption.

The precise relation between the correlators $\omega_{g,n}$ and the linear Hodge integrals (or Hurwitz numbers) is given by:
\begin{theorem}[\cite{Eynard2009TheLT}]\label{Thm:Lambert}
    Let the Lambert spectral curve be $(\mathbb{P}^1\setminus \mathbb{R}_-,x(z)=-z+\log(z),y(z)=\log(z), \frac{dz_1\,dz_2}{(z_1-z_2)^2})$, then the correlator $\omega_{g,n}$ computed by TR  generates the linear Hodge integrals
    \begin{align}\label{LambertCurve}
        \omega_{g,n}(z_1,...,z_n)=\sum_{k_1,...,k_n\geq 0}\prod_{i=1}^n\frac{k_i^{k_i+1}}{k_i!} \bigg\langle\frac{\Lambda(1)}{\prod_{i=1}^n(1-k_i\psi_i)}\bigg\rangle_{g,n}e^{k_i x_i(z_i)} dx_i(z_i)
    \end{align}
\end{theorem}
Computing the Laplace transform of \eqref{LambertCurve}, where $\gamma$ is a contour encircling the origin, will separate the summands in  \eqref{LambertCurve}:
\begin{lemma}\label{lem:lambert}
Let $\mu_i\in \mathbb{N}$ and $2g+n-2>0$, we have for $\omega_{g,n}$ generated by the Lambert curve of Theorem \ref{Thm:Lambert}
\begin{align}\label{eq:lambert}
    \Res_{z_i=0} e^{-\mu_1x_1(z_1)...-\mu_nx_n(z_n)}\omega_{g,n}(z_1,...,z_n)=\prod_{i=1}^n\frac{\mu_i^{\mu_i+1}}{\mu_i!} \bigg\langle\frac{\Lambda(1)}{\prod_{i=1}^n(1-\mu_i\psi_i)}\bigg\rangle_{g,n}.
\end{align}
For $2g+n-2\leq0$, we take that the lhs of \eqref{eq:lambert} to be defined by this equation with $\bigg\langle\frac{\Lambda(1)}{(1-\mu\psi)}\bigg\rangle_{0,1}=\frac{1}{\mu^2}$ and $\bigg\langle\frac{\Lambda(1)}{(1-\mu_1\psi_1)(1-\mu_2\psi_2)}\bigg\rangle_{0,2}=\frac{1}{\mu_1+\mu_2}$.
\begin{proof}
    For $2g+n-2>0$ and each $\mu_i$, we have the same computation
    \begin{align*}
        \Res_{z=0} dx(z) e^{(k-\mu)x(z)}=\Res_{z=0} \frac{(1-z)dz}{z} \frac{e^{(\mu-k)z}}{z^{\mu-k}}=\delta_{\mu,k}.
    \end{align*}
    For $2g+n-2<0$, the residue at 0 is not well-defined. Thus, we define the Laplace transform to match 
    the classical known result.
    The case $(g,n)=(0,1)$ is related to Remark \ref{rem:01}.
\end{proof}
\end{lemma}

Now, we can apply the formulas developed in Sec. \ref{sec:laplace} to the Lambert curve. First of all note that the Lambert curve parametrised by $x(z)=-z+\log(z),y(z)=\log(z)$, i.e.
\begin{align}\label{LambertCurve2}
    x=-e^y+y,
\end{align}
is not ramified through $y$. We can easily stick to the examples discussed in Sec. \ref{sec:laplace} and just have to compute the following primitives:
\begin{align}\label{Phi1L}
    \Phi_1^\vee(y(z))&=\frac{1}{\hbar}\int x(z)dy(z)=\frac{-z+\frac{\log(z)^2}{2}}{\hbar}=\frac{-e^{y(z)}+\frac{y(z)^2}{2}}{\hbar}\\\label{Phi1HL}
    \hat{\Phi}^\vee_1(y(z);\hbar,u)&=\Phi_1^\vee\bigg(y(z)+\frac{\hbar u}{2}\bigg)-\Phi_1^\vee\bigg(y(z)-\frac{\hbar u}{2}\bigg)=\\\nonumber
    &=-z u S(\hbar u)+\log(z) u\\\label{Phi2L}
    \Phi_2^\vee(y_1(z_1),y_2(z_2))&=\log(z_1-z_2)\\\label{Phi1HL2}
    \hat{\Phi}_2^\vee(y_1(z_1),y_2(z_2);\hbar,u_1,u_2)&=\log\bigg(\frac{(z_1e^{\hbar u_1/2}-z_2e^{\hbar u_2/2})(z_1e^{-\hbar u_1/2}-z_2e^{-\hbar u_2/2})}{(z_1e^{-\hbar u_1/2}-z_2e^{\hbar u_2/2})(z_1e^{\hbar u_1/2}-z_2e^{-\hbar u_2/2})}\bigg)\\\label{Phi2LD}
    \hat{\Phi}_2^\vee(y(z),y(z);\hbar,u,u)&=-2\log(S(u\hbar)),
\end{align}
where $S(u)=\frac{e^{u/2}-e^{-u/2}}{u}$ and $\Phi_2^\vee$ needed to be regularised due to \eqref{PhiHutDia} on the diagonal.
\begin{remark}
    Note that we have chosen $y(z)=\log(z)$ instead of $y(z)=z$ as proposed in \cite{Bouchard:2007hi}, because Sec. \ref{sec:logxy} has shown that for  $y(z)=z$ the functional relation does not hold. Actually, transforming $y(z)=z\mapsto y(z)+x(z)=\log(z)$, is the first symplectic transformation of a spectral curve mentioned in the beginning of Sec. \ref{sec:laplace}. Thus, both curves have the same $\omega_{g,n}$'s, except for $\omega_{0,1}$, and therefore generates the linear Hodge integrals as stated in Theorem \ref{Thm:Lambert}. The important factor $S(u\hbar)$ appearing for Hurwitz numbers and linear Hodge integrals (see for instance \cite{Faber:1998gsw}) is directly produced in \eqref{Phi1HL} and \eqref{Phi2LD} by our chosen spectral curve \eqref{LambertCurve2} and the formula of $x-y$ symplectic transformation.
\end{remark}

Taking Example \ref{Ex:4} and Lemma \ref{lem:lambert} into account, we conclude:
\begin{corollary2}\label{cor:lambert}
    The linear Hodge integrals are computed as a formal expansion in $\hbar$ via
    \begin{align*}
        &\prod_{i=1}^n\frac{k_i^{k_i+1}}{k_i!}\bigg\langle\frac{\Lambda(1)}{\prod_{i=1}^n(1-k_i\psi_i)}\bigg\rangle_{g,n}\\
        =&\Res_{z_i=0}[\hbar^{2g+n-2}]\prod_{i=1}^n\frac{dz_i e^{k_iz_iS(\hbar k_i)}}{z_i^{1+k_i}\hbar k_i S(\hbar k_i) }\\
        &\qquad \qquad \times \sum_{\Gamma\in\tilde{\G}^2_n}\prod_{\substack{I\in\I(\Gamma)\\\{i,j\}=I}}\bigg(\frac{(z_ie^{\hbar k_i/2}-z_je^{\hbar k_j/2})(z_ie^{-\hbar k_i/2}-z_je^{-\hbar k_j/2})}{(z_ie^{-\hbar k_i/2}-z_je^{\hbar k_j/2})(z_ie^{\hbar k_i/2}-z_je^{-\hbar k_j/2})}-1\bigg),
    \end{align*}
    where $\tilde{\G}^2_n\subset \G_n$ is the subset of graphs with 2-valent $\bullet$-vertices connecting two different $\bigcirc$-vertices at most with one $\bullet$-vertex which is nothing than the set of connected graphs with $n$ labelled vertices (\href{https://oeis.org/A001187}{A001187}).
\end{corollary2}
Explicit computations of Corollary \ref{cor:lambert} with computer algebra have matched all the Hurwitz numbers listed in \cite[Tab. 2]{Eynard2009TheLT}.

For not-necessarily-connected correlators, we can conclude from Example \ref{Ex:3} and Lemma \ref{lem:lambert}:
\begin{corollary2}
    The sum over all partitions of linear Hodge integrals satisfies as a formal expansion in $\hbar$
    \begin{align*}
        &\prod_{i=1}^n\frac{k_i^{k_i+1}}{k_i!}\sum_{\lambda\vdash I}\prod_{j=1}^{l(\lambda)}\sum_{g_j=0}\hbar^{2g_j+|\lambda_j|-2} \bigg\langle\frac{\Lambda(1)}{\prod_{i=1}^{|\lambda_j|}(1-k_{\lambda_j^i}\psi_{\lambda_j^i})}\bigg\rangle_{g_j,|\lambda_j|}\\
        =&\prod_{i=1}^n\Res_{z_i=0}\frac{dz_i e^{k_iz_iS(\hbar k_i)}}{z_i^{1+k_i}\hbar k_i S(\hbar k_i) }
         \prod_{i<j}\frac{(z_ie^{\hbar k_i/2}-z_je^{\hbar k_j/2})(z_ie^{-\hbar k_i/2}-z_je^{-\hbar k_j/2})}{(z_ie^{-\hbar k_i/2}-z_je^{\hbar k_j/2})(z_ie^{\hbar k_i/2}-z_je^{-\hbar k_j/2})},
    \end{align*}
    where $\lambda\vdash I$ is a set partition of $I$ with $l(\lambda)$ blocks $\lambda_j\subset I$, i.e. $\lambda=(\lambda_1,...,\lambda_{l(\lambda)})$. Each block is written as $\lambda_j=(\lambda_j^1,...,\lambda_j^{|\lambda_j|})$ of cardinality $|\lambda_j|$ and elements $\lambda_j^i\in I$.
\end{corollary2}

\bibliographystyle{halpha-abbrv}
\bibliography{omega.bib}
\end{document}